\documentclass[prd,twocolumn,preprintnumbers,showpacs,nofootinbib,superscriptaddress,notitlepage,floatfix]{revtex4-1}
\usepackage{hyperref}
\usepackage{graphicx}
\usepackage{amsmath}
\usepackage{amssymb}
\usepackage{slashed}
\usepackage{amsfonts}
\usepackage{xcolor}
\usepackage{multirow}
\usepackage[caption=false]{subfig}
\usepackage{array}
\usepackage{mwe}
\usepackage{dsfont}
\usepackage{soul}
\newcommand{\be}{\begin{equation}}
\newcommand{\ee}{\end{equation}}
\newcommand{\bea}{\begin{eqnarray}}
\newcommand{\eea}{\end{eqnarray}}
\newcommand{\ba}{\begin{array}}
\newcommand{\ea}{\end{array}}

\usepackage[utf8]{inputenc}
\usepackage[normalem]{ulem}

\begin{document}
\preprint{LA-UR-24-26903}
%%%%%%%%%%%%%%%%%%%%%%%%%%%%%%%%%%%%%%%%%%%%%%%%%%%%%%%%%%%%%%%%%%%%%%
\title{Threshold photo-production of $J/\Psi$ off light nuclei}
%%%%%%%%%%%%%%%%%%%%%%%%%%%%%%%%%%%%%%%%%%%%%%%%%%%%%%%%%%%%%%%%%%%%%%

\author{Fangcheng He}
\email{fangchenghe@lanl.gov}
\affiliation{Center for Nuclear Theory, Department of Physics and Astronomy, Stony Brook University, Stony Brook, New York 11794–3800, USA}
\affiliation{Los Alamos National Laboratory, Theoretical Division T-2, Los Alamos, NM 87545, USA}
\author{Ismail Zahed}
\email{ismail.zahed@stonybrook.edu}
\affiliation{Center for Nuclear Theory, Department of Physics and Astronomy, Stony Brook University, Stony Brook, New York 11794–3800, USA}

\date{\today}

\begin{abstract}
We analyze threshold photoproduction of heavy mesons off a deuteron and Helium-4, using the QCD factorization method. Assuming large skewness,
the production amplitude is dominated by the leading twist-2 gluonic energy-momentum  tensor (EMT). We use our recent results for the gluonic gravitational form factors of light nuclei in the impulse approximation, to estimate the differential cross sections for $J/\Psi$ production off a deuteron and Helium-4 at current electron facilities.
\end{abstract}

\maketitle

{\bf 1. Introduction\,\,}
 Gluons play a central  role in our understanding of the QCD vacuum structure, and the formation of hadronic states~\cite{Schafer:1996wv,Biddle:2019gke}. Their non-perturbative and topological character at low resolution, is at the origin of the breaking of conformal and chiral symmetries in QCD,  the emergence of mass from  no mass~\cite{Zahed:2021fxk} (and references therein).
Unlike the quarks, the gluons are not electrically charged and therefore difficult to probe directly  using current electron machines.

For sufficiently high energy, gluons can be probed in the form of jets. Their fragmentation into hadrons may provide some insights on their role in the composition of hadrons.  Alternatively, coherent threshold electro- or photo-production of heavy mesons off hadrons, is sensitive to gluon exchanges, a way to probe the 
gluonic content of hadrons at lower resolution. The recent experiments 
 carried at JLAB~\cite{GlueX:2019mkq,Meziani:2020oks,Duran:2022xag} have started to reveal some aspects of the  gluon substructure in the proton. More experiments along these lines are planned at the future electron ion collider (EIC).
 
 Near threshold diffractive  electro- or photo-production of  heavy mesons such as charmonium or bottomonium, is sensitive to the gluon content of the probed hadron~\cite{Burkert:2022hjz} (and references therein).   The JLAB results have increased considerably the interest in this process, in light of the fact that they may directly probe the proton gluonic gravitational form factors~\cite{Hatta:2018ina,Mamo:2019mka,Kharzeev:2021qkd,Ji:2021mtz,Hatta:2021can,Guo:2021ibg,Sun:2021gmi,Mamo:2022eui,Wang:2022vhr,Guo:2023pqw}.
That gluons dominate the diffractive vector meson production at large center of mass energy $\sqrt s$ is
not surprising~\cite{Nemchik:1996cw}. What is  surprising, is that they may still dominate the threshold production of heavy quarkonia. Indeed, at large $\sqrt s$ diffractive $pp$  and $p\bar p$ is dominated by Pomeron exchange a tower a C-even soft gluons with positive signature, with a small Odderon admixture, a tower of C-odd soft gluons with negative signature~\cite{TOTEM:2020zzr}.  Negative signature  Reggeons add in the $pp$ channel, and subtract in the $p\bar p$ channel,
as suggested by the recent TOTEM data at LHC~\cite{TOTEM:2018hki,TOTEM:2020zzr}.

Threshold $J/\Psi$ photo-production at JLAB has opened the possibility of measuring  the gluonic gravitational form factors of the proton~\cite{GlueX:2019mkq,Meziani:2020oks}.
The results have been analyzed using QCD factorization~\cite{Guo:2021ibg,Guo:2023pqw} and dual gravity~\cite{Mamo:2019mka,Mamo:2021krl}, both with a fair  account of the reported data. In the former, the near threshold amplitude is factorized using generalized parton distributions (GPDs), and shown to be dominated by the twist-2 part of the energy-momentum tensor. In the latter, the dual amplitude is dominated by the exchange of a graviton (traceless plus tracefull) in bulk which directly maps onto
the energy-momentum tensor at the boundary. Dual gravity yields explicit gravitational form factors (GFFs)~\cite{Abidin:2009hr,Mamo:2019mka,Mamo:2021krl}. They  carry important information on the nucleon mass, angular momentum, pressure and shear force.

Threshold photo-production of heavy mesons off light nuclei such as a deuteron or Helium-4, if detectable at JLAB or future facilities such as the EIC, may provide for further understanding of how the gluonic exchanges get redistributed in few nucleon systems, in the presence of meson exchanges. It may also allow for the possibility of extracting the meson GFFs through selecting exchange currents, in analogy with electron scattering on light nuclei. The purpose of this letter is to address the threshold photo-production on light nuclei following the QCD factorization method~\cite{Guo:2021ibg,Guo:2023pqw}. The  gravity dual approach will be presented elsewhere. For completeness, we note the recent proposal for the electro-production process at the EIC, based on gluon shadowing by few nucleons~\cite{Guzey:2022jtv}.

\vskip 1cm
{\bf 2. Photo-production on light nuclei\,\,}
Threshold  photo-production of a heavy meson on a nucleon using the QCD factorization method, has been used recently for charmonium in~\cite{Guo:2021ibg,Guo:2023pqw} and
for $\eta_c$ in~\cite{Liu:2024yqa}.
In short, the threshold amplitude is factorized into a hard kernel times a  gluon GPD. In the heavy meson limit, the GPD is dominated by the leading moments, which are tied to the gluonic GFFs in the threshold region. For large skewness, the gluonic GPD is  dominated 
by the leading twist-2 gluon gravitational form factors.

The QCD factorization method can be extended to coherent $J/\Psi$ production on light nuclei near threshold, essentially with the same  assumptions. The leading twist-2 
gravitational form factors are those of light nuclei we have
recently derived in~\cite{He:2023ogg,He:2024vzz,He:2024jgc}.
More specifically, in the photo-production process, the flowing partons
(gluons) carry large momenta $k^+_i$ times the pertinent  parton correlation in a generic light nuclear target  $N=$ deuteron, Helium-4.  The result is~\cite{Guo:2021ibg,Guo:2023pqw}
\begin{widetext}
\bea
    i\mathcal{M}(\gamma N\rightarrow J/\psi\ N')
    =&&\int_{-\bar{P}^+}^{\infty} dk^+_1\int_{-\bar{P}^+}^{\infty}dk^+_2W^{ab}_{\mu\nu}(k_1^+,k_2^+)\frac{i}{k_1^++i0^+}\frac{i}{k_2^++i0^+}\nonumber\\
    &&\times \int \frac{d\lambda_1^-}{2\pi}\frac{d\lambda_2^-}{2\pi}e^{-ik^+_1\lambda_1^--ik^+_2\lambda_2^-}\langle N'|F^{a\mu+}\left(\lambda_2^-\right) F^{b\nu+}\left(\lambda_1^-\right)|N\rangle
\eea
\end{widetext}
where the  symmetric parameterization is assumed. One can define
$
\lambda_c^-={(\lambda_1^-+\lambda_2^-)}/{2}, \Delta^+=k^+_1+k_2^+=-2\xi\bar{P}^+, 
\lambda^-=\lambda_1^--\lambda_2^-, k^+={(k^+_1-k_2^+)}/{2}=x\bar{P}^+
$.
Translational symmetry causes $\lambda_c^-$ to drop from the  partonic correlator.
The lower bound  ensures that the spectator parton in $N$ is physical.  $\Delta^+$ drops out by translational symmetry. For $X=J/\Psi, \Upsilon$, the transverse polarization 
dominates the amplitude in the heavy quark limit~\cite{Sun:2021pyw} 
\bea
    W^{ab}_{\mu\nu}=\frac{g^2}{2}\frac{\delta^{ab}g_{\perp\mu\nu}}{\sqrt{N_c}}\frac{\psi^*_{X}(0)}{\sqrt{m_V^3}}(8\epsilon_\gamma\cdot\epsilon_X^*)
\eea
Here  $\psi_{X}$ is the non-relativistic wave function for quarkonium. In the heavy meson photoproduction process, the relative momentum between the quark and antiquark is of order $\mathcal{O}(\alpha_sM_X)$, and the heavy meson mass is assumed to be $M_X=2m_Q$.

The factorized amplitude  is derived by the leading twist-2 
GPD for the gluonic energy momentum tensor on the light cone~\cite{Sun:2021pyw} 
%\begin{widetext}
\begin{equation}
\begin{aligned}
    &i\mathcal{M}(\gamma N\rightarrow X\ N')= \frac{g^2}{\sqrt{N_c}}\frac{\psi^*_{J/\psi}(0)}{\sqrt{m_V^3}}(4\epsilon_\gamma\cdot\epsilon_V^*)\mathcal{W}_{2g}(t,\xi)
\end{aligned}
\end{equation}
with $\mathcal{W}_{2g}(t,\xi)$ defined as
\begin{equation}
\label{GPD0}
    \mathcal{W}_{2g}(t,\xi)=\int_{-1}^{1} dx\frac{1}{x-\xi+i0^+}\frac{1}{x+\xi-i0^+}f_{2g}(x,t,\xi)
\end{equation}
and with the  gluonic GPD 
\begin{widetext}
\begin{equation}
\label{GPD1}
    f_{2g}(x,\xi,t)=\int \frac{d\lambda^-}{2\pi}e^{-ix\bar{P}^+\lambda^-}\frac{1}{\bar{P}^+}\langle P'|F^{a+i}\left(-\lambda^-/2\right) F^{a+}{}_i\left(\lambda^-/2\right)|P\rangle
\end{equation}

For large skewness, the dominant contribution stems from the leading
local bilinear
$F^{a+i}\left(-\lambda^-/2\right) F^{a+}{}_i\left(\lambda^-/2\right)\simeq F^{a+i}F^{a+}{}_i(0)$,
which once inserted in (\ref{GPD0}) gives the off-forward matrix element of the gluonic energy-momentum tensor in a light nuclear target
\bea
\label{GPD2}
\mathcal{W}_{2g}(t,\xi\rightarrow1)=-\frac{1}{\xi^2(\bar{P}^+)^2}\langle P'|F^{a+i} F^{a+}{}_i|P\rangle\equiv -\frac{1}{\xi^2(\bar{P}^+)^2}\langle P'|T^{++}_g|P\rangle
\eea

\vskip 1cm
{\bf 3. Deuteron, Helium-4\,\,}
The simplest light nuclear target is Helium-4, with a single gluonic distribution $H_g(x,t)$, 
%\cite{Ji:1998pc}. 
\begin{equation}
\label{GPD1HE}
    f_{2g}(x,\xi,t)=\int \frac{d\lambda^-}{2\pi}e^{-ix\bar{P}^+\lambda^-}\frac{1}{\bar{P}^+}\langle H',P'|F^{a+i}\left(-\lambda^-/2\right) F^{a+}{}_i\left(\lambda^-/2\right)|H, P\rangle \equiv H_g(x,\xi, t)
\end{equation}
The zeroth moment of the gluon distribution dominates in the threshold region. It is tied to the energy momentum tensor in a Helium-4 target
\bea
H_{2g}(\xi, t)=\int_{-1}^{1}dx\,H_g(x,\xi, t)=
 \frac{1}{(\bar{P}^+)^2}\langle H',P'|T^{++}_g|H,P\rangle=
2A_g^H(t)+2\xi^2\,D^H_g(t)
\eea
\end{widetext}
with $H_{2g}(-\xi, t)=H_{2g}(\xi, t)$ by time-reversal symmetry.  The   gluonic gravitational form factors for Helium-4 are defined as
\bea
\label{THE4}
\langle H',P'|T^{++}_g|H,P\rangle=2{\bar P}^{+2}A^H_g+
\frac{{\Delta^{+2}}}{2}\,D^H_g(t)
\eea
Similarly, the covariant EMT matrix element of the deuteron is defined as~\cite{Polyakov:2019lbq}
\begin{widetext}
\bea\label{eq:TppD}
\langle P',m'|T^{++}_g|P,m\rangle &=&
2(\bar{P}^+)^2   \left[
- {\epsilon^{\prime*}\cdot \epsilon} \, A^g_0 (t) 
 +{ {\epsilon^{\prime*}\cdot \bar{P}} \, {\epsilon \cdot \bar{P}} \over m_D^2}
 \, A^g_1(t) \right]
+\frac12( \Delta^+)^2
 \left[
{\epsilon^{\prime*}\cdot \epsilon} \, D^q_0 (t)
+{ {\epsilon^{\prime*}\cdot \bar{P}} \, {\epsilon \cdot \bar{P}} \over m_D^2}  \, D^g_1(t)\right]
\nonumber\\
&&+4\bar{P}^+  \left[\epsilon^{\prime*+} \,\epsilon\cdot \bar{P}+\epsilon^{+}\,
\epsilon^{\prime*}\cdot \bar{P}\right] \, J^g (t)+\Bigl[ \epsilon^{+}
\epsilon^{\prime*+} \Delta^2
-2\epsilon^{\prime*+}\Delta^+ \,\epsilon\cdot \bar{P}  
+2\epsilon^{+} \Delta^+  \,\epsilon^{\prime*}\cdot \bar{P}
  \Bigl] \, E^g(t)
\nonumber\\
\eea
\end{widetext}
The corresponding covariant form factors \{$A^g_0 (t)$, $A^g_1(t)$, $D^g_0(t)$, $D^g_1 (t)$, $J^g(t)$, $E^g(t)$\} can be expressed by the form factors defined in the Breit frame \{$\mathcal A^g(t)$, $\mathcal Q^g(t)$, $\mathcal J^g(t)$, $\mathcal D^g_0 (t)$, $\mathcal D^g_2(t)$, $\mathcal D^g_3(t)$\} using the relations shown in~\cite{Polyakov:2019lbq}
\begin{widetext}
\bea\label{eq:tranf}
A^g_0(t) &=& \frac{12 \mathcal A^g(t) m_D^2+3 \mathcal D^g_0 (t) t+4 \mathcal D^g_2(t) t+\mathcal D^g_3(t) t-2 \mathcal Q^g (t) t}{3 \left(4 m_D^2-t\right)},
\nonumber\\
D^g_0(t) &=& -\frac{1}{3} (3 \mathcal D^g_0 (t)+4 \mathcal D^g_2(t)+\mathcal D^g_3(t))
\nonumber\\
J^g(t) &=& \frac{\mathcal D^g_2(t) t+4 \mathcal J^g(t) m_D^2}{4 m_D^2-t}
\nonumber\\
E^g(t) &=& -\mathcal D^g_2(t)
\nonumber\\
A^g_1(t)&=& \frac{8 m_D^2 }{3 \left(t-4 m_D^2\right)^2}\Big[12 \mathcal A^g(t) m_D^2+3 \mathcal D^g_0 (t) t+\mathcal D^g_2 (t) \left(t-12 m_D^2\right)-6 \mathcal D^g_3 (t) m_D^2+\mathcal D^g_3 (t) t
\nonumber\\
&-&24 \mathcal J^g(t) m_D^2+12 m_D^2 \mathcal Q_g(t)-2 \mathcal Q_g(t) t\Big]
\nonumber\\
D^g_1(t) &=& \frac{8 m_D^2 \left(3 \mathcal D^g_0 (t) t+\mathcal D^g_2(t) t-6 \mathcal D^g_3(t) m_D^2+\mathcal D^g_3(t) t\right)}{3 t \left(4 m_D^2-t\right)}
\eea
\end{widetext}
For convenience, the GFFs in the Breit frame are summarized in~\ref{eq:GFFBre}. The  latters have been recently analyzed in the impulse approximation~\cite{He:2023ogg}, and including the exchange current corrections  in~\cite{He:2024vzz}. The exchange corrections 
for Helium-4 were found to be very small, especially when the  pseudoscalar nucleon-pion coupling was used~\cite{He:2024jgc}.
Whence, we will limit our discussion of the photo-production process to the impulse approximation.

\begin{figure*}
\begin{center}
\includegraphics[scale=0.6]{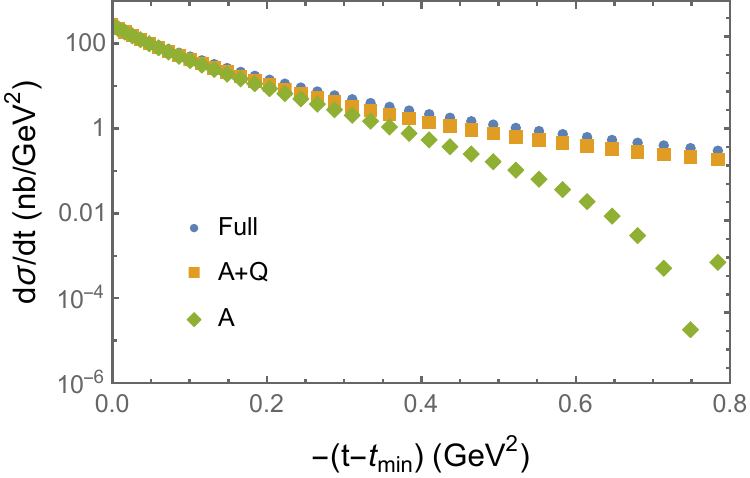}
\includegraphics[scale=0.6]{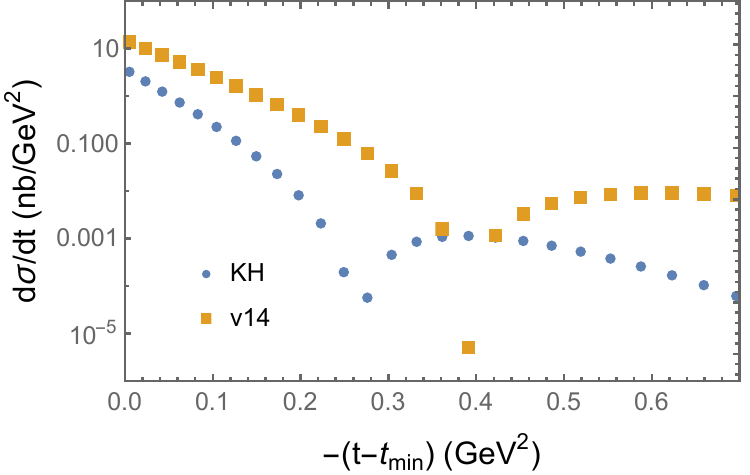}
\caption{The photo-production differential cross section for the  deuteron (left panel) and Helium-4 (right panel) using the impulse approximation with  $W=10$ GeV. ``KH" and ``v14" represent the results obtained using the GFFs with the K-Harmonic method and the Agronne v14 potential~\cite{He:2023ogg,He:2024vzz}.}
\label{fig:DH4}
\end{center}
\end{figure*}

\vskip 1cm
{\bf 4. Differential cross section\,\,}
\label{SEC4}
%\subsection{Helium-4}
The differential cross section for threshold photo-production of $X=J/\psi$ on spin targets, follows from standard arguments.
\begin{equation}
  \frac{d\sigma}{dt}=\frac{Q_c^2e^2}{16\pi(s-M_N^2)^2}\frac 12\sum_{\mathrm{polarizations}}|\mathcal{M}|^2  
\end{equation}
with the kinematics detailed in appendix~\ref{KINEMATICS}.
For Helium-4 with spin-0, the differential cross section
yields
%For Helium-4, we have
\begin{widetext}
\bea
\label{DSIGT}
\left(\frac{d\sigma}{dt}\right)_{He}&=& 4\pi\alpha_{em}Q_c^2\frac{16\pi\alpha^2_s}{{4}(s-M_N^2)^2}\frac{4}{N_cM^3_{J/\psi}}\left|\psi_{J/\psi}(0)\right|^2
\frac{1}{2}\sum_{\lambda_\gamma \lambda_V}(\epsilon_\gamma\cdot\epsilon_V^*)^2\left|\mathcal{W}_{2g}(t,\xi)\right|^2\nonumber\\
&=&4\pi \alpha_{em}Q_c^2\frac{16\pi\alpha^2_s}{{4}(s-M_N^2)^2}\frac{4}{N_cM^3_{J/\psi}}\left|\psi_{J/\psi}(0)\right|^2
\frac{4}{\xi^4}
\bigg(A_g^H(t)+{\xi^2}\,D^H_g(t)\bigg)^2   
\eea
\end{widetext}
and for the deuteron with spin-1, we have
\begin{widetext}
\bea
\label{DSIGT}
\left(\frac{d\sigma}{dt}\right)_D&=& 4\pi\alpha_{em}Q_c^2\frac{16\pi\alpha^2_s}{{4}(s-M_N^2)^2}\frac{4}{N_cM^3_{J/\psi}}\left|\psi_{J/\psi}(0)\right|^2
\frac{1}{2}\sum_{\lambda_\gamma \lambda_V}(\epsilon_\gamma\cdot\epsilon_V^*)^2
\sum_{\epsilon'\epsilon}\frac13\left|\mathcal{W}_{2g}(t,\xi)\right|^2\\
&=&4\pi \alpha_{em}Q_c^2\frac{16\pi\alpha^2_s}{{4}(s-M_N^2)^2}\frac{4}{N_cM^3_{J/\psi}}\left|\psi_{J/\psi}(0)\right|^2\frac{1}{9 \xi^4\left(t-4 m_D^2\right)^2 }
\nonumber\\
&\times&\Bigg\{
4 \Big[144 (\mathcal A^g)^2 m_D^4+72 \mathcal A^g \mathcal D^g_0  m_D^2 t+t \left(t \left(9 (\mathcal D^g_0)^2+8 (\mathcal D^g_2)^2+4 \mathcal D^g_2  (\mathcal D^g_3 -2 \mathcal Q^g)+2 (\mathcal D^g_3-2 \mathcal Q^g)^2\right)-96 (\mathcal J^g)^2 m_D^2\right)\Big]
\nonumber\\
&+& 8\xi^2(4m_D^2-t) \Big[36 \mathcal A^g \mathcal D^g_0  m_D^2+t \left(9 (\mathcal D^g_0)^2+8 (\mathcal D^g_2)^2+4 \mathcal D^g_2 (\mathcal D^g_3 -\mathcal Q^g)+2 \mathcal D^g_3 (\mathcal D^g_3 -2 \mathcal Q^g)\right)-48 (\mathcal J^g)^2 m_D^2\Big]
\nonumber\\
&+&4\xi ^4 \left(t-4 m_D^2\right)^2\Big[9 (\mathcal D^g_0)^2+8 (\mathcal D^g_2)^2+4 \mathcal D^g_2 \mathcal D^g_3 +2 (\mathcal D^g_3)^2\Big]
\Bigg\}
\eea
\end{widetext}
We made use of~(\ref{GPD2}), (\ref{eq:TppD}-\ref{eq:tranf}) and the polarization sum rule for the spin one target $$\sum_\lambda\epsilon^{*\mu}(P,\lambda)\epsilon^{\nu}(P,\lambda)=-g^{\mu\nu}+\frac{P^\mu P^\nu}{m_D^2}\,.$$
and only the transverse polarizations were retained for the heavy mesons~\cite{Sun:2021pyw,Guo:2021ibg}. The  differential cross section for a spin-averaged deuterium target follows similarly.
To proceed numerically, we set the charmonium mass to
$M_{J/\Psi}= 3.097\,{\rm GeV}$~\cite{ParticleDataGroup:2018ovx}. 
The strong coupling constant at this scale is
$\alpha_s(2m_c)\approx 0.31$.
The heavy meson wavefunction at the origin is fixed by the decay constant~\cite{Bodwin:2006yd}
\begin{equation}
    \Gamma(J/\psi\rightarrow e^+e^-)=\frac{16\pi\alpha_{em}^2Q_c^2}{3M^2_{J/\psi}}N_c|\psi_{J/\psi}(0)|^2\left(1-\frac{16}{3}\frac{\alpha_s}{\pi}\right)
\end{equation}
with the empirical value of $5.55\,{\rm keV}$~\cite{ParticleDataGroup:2018ovx}, hence $|\psi_{J/\psi}(0)|^2= 0.094\,{\rm GeV}^3$.

In Fig.~\ref{fig:DH4} (left) we show the results for the  threshold photo-production of $J/\Psi$ on a deuteron  target at $W=10$ GeV. The diffractive dip generated by the mass GFF ${\cal A}^g$  (green diamond), is washed out by the addition of the quadrupole GFF  ${\cal Q}^g$ (orange squares). Both GFFs are assessed in the impulse approximation using the results in~\cite{He:2023ogg}. This suggests that in the dip region, the quadrupole $Q$ GFF is potentially measureable. 

In Fig.~\ref{fig:DH4} (right)
 we show the results for the  threshold photo-production of $J/\Psi$ on a Helium-4 target also at $W=10$ GeV. In contrast to the deuteron, the diffractive dip is noticeable in the dipole approximation for both the K-harmonic (blue circles) and the Argonne v14 potential (orange diamond)~\cite{He:2024jgc} with D-wave admixture.

\begin{figure*}
\subfloat[\label{fig:WtheD}]{%
\includegraphics[height=5cm,width=.45\linewidth]{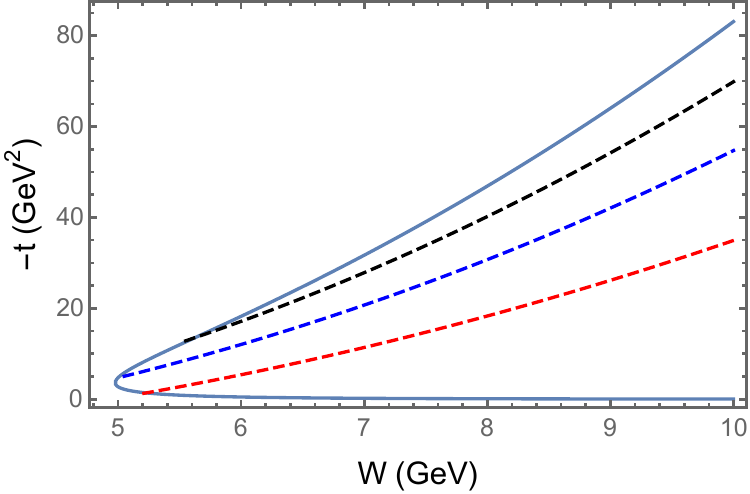}%
}\hfill
\subfloat[\label{fig:Wthelium4}]{%
\includegraphics[height=5cm,width=.45\linewidth]{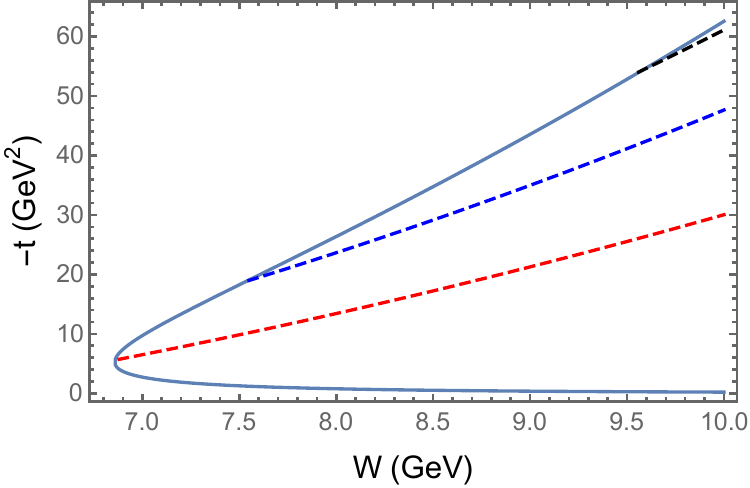}%
}
\caption{ a: The allowed region for $W$ and t for the deuteron. The red, blue and black dashed lines represent the results with $\xi=0.3\,, 0.5\,,0.7$ respectively;
b:The allowed region for $W$ and t for Helium-4, with the same color coding.}
\label{fig:KIN}
\end{figure*}

We note that the value of the center of mass energy used  $W=10\, \rm GeV$ implies a low value of the skewness parameter from Fig.~\ref{fig:KIN}, suggesting that higher corrections in $\xi$ to (\ref{DSIGT}) maybe needed. However, we recall that for the nucleon case, dual gravity arrives at a similar result (in the large $N_c$ limit) without assuming large skewness~\cite{Mamo:2021krl}.
%to the skewness expansion, nor to factorization.

\vskip 1cm
{\bf 5. Conclusions\,\,}
Threshold coherent photo-production of heavy mesons at current and future electron facilities, has the potential of probing the gluonic content of the nucleon at low resolution. The recent JLAB measurements of $J/\Psi$ off nucleon targets~\cite{GlueX:2019mkq,GlueX:2023pev,Duran:2022xag}, have provided the first detailed differential  cross sections in the near threshold region, in fair agreement with  the predictions from  QCD factorization~\cite{Guo:2021ibg,Guo:2023pqw}and dual gravity~\cite{Mamo:2019mka,Hatta:2021can,Mamo:2022eui}.

We have now extended the QCD factorization method to 
the coherent photo-production of $J/\Psi$ off light nuclei near threshold, using the GFFs recently derived in~\cite{He:2023ogg}. 
The empirical results for the differential cross sections, can be used to extract the gluonic GFFs and radii of these light nuclei. We look forward to their possible measurements currently at JLAB, and in the near future at the EIC.

\vskip 1cm
{\bf Acknowledgments\,\,}
We thank Sangbaek Lee and Zein-Eddine Meziani for discussions. F.H. acknowledges discussions with Tanmoy Bhattacharya and Rajan Gupta who were partly supported by the U.S. Department of Energy, Office of Science, Office of High Energy Physics under Contract No. DE-AC52-06NA25396. F.H. is supported by the National Science Foundation under CAREER Award PHY-1847893 and the LANL Laboratory Directed Research and Development (LDRD) program. I.Z. is supported by the Office of Science, U.S. Department of Energy under Contract  No. DE-FG88ER40388. This research is also supported in part within the framework of the Quark-Gluon Tomography (QGT) Topical Collaboration, under contract no. DE-SC0023646.

\appendix

\section{Kinematics}
\label{KINEMATICS}
The relevant kinematic invariants for the meson photoproduction are Mandelstam $s$ and $t$. $s=(P+q)^2$ is related to the center of mass energy $W=\sqrt{s}$ and $t=\Delta^2$ is related to the momentum transfer $\Delta^\mu=(P'-P)^\mu$. Different from the leptoproduction, the $Q^2$ in photoproduction is exactly set to be $0$ although similar analysis can be easily extended for the large-$Q^2$ leptoproduction.
Without loss of generality, we can work in the center of mass frame. The four-momenta of the incoming photon, incoming proton, outgoing proton and outgoing meson $X$ are denoted by $q$, $P$, $P'$, and $q'$ respectively. Each external state is given by the on-shell conditions defined as $P^2=P'^2=M_N^2\,, q^2=0\,,  q'^2=M_X^2$.
With the on-shell conditions, the four-momenta in the center of mass frame, can be written as
\begin{align}
\label{eq:kinematics}
q=&\left(\frac{s-M^2_N}{2\sqrt{s}},\ 0,\ -\frac{s-M^2_N}{2\sqrt{s}}\right)  & \\[5pt] \nonumber 
q'=&\left(\frac{s+M_X^2-M^2_N}{2\sqrt{s}},\ -|\vec{P}'_{c}|\sin\theta,\ -|\vec{P}'_{c}|\cos\theta\right) \\[5pt] \nonumber
P=&\left(\frac{s+M^2_N}{2\sqrt{s}},\ 0,\ \frac{s-M^2_N}{2\sqrt{s}}\right)  &  \\[5pt] \nonumber
P'=&\left(\frac{s-M_X^2+M^2_N}{2\sqrt{s}},|\vec{P}'_{c}|\sin\theta,\ |\vec{P}'_{c}|\cos\theta \right)
\end{align}
where $M_N$ is the mass of the light nuclei $N=$deuteron, Helium-4, $M_X$ is the produced meson mass, and $\theta$ is the scattering angle in the center of mass frame. The magnitude of the outgoing three-momentum reads
\begin{equation}
|\vec{P}'_c|=\left(\frac{[s-(M_X+M_N)^2][s-(M_X-M_N)^2]}{4s}\right)^{1/2}    
\end{equation}   
The scattering angle is determined by the invariant $t$
\begin{equation}
    \cos\theta=\frac{2st+(s-M^2_N)^2-M^2_X(s+M_N^2)}{2\sqrt{s}|\vec{P}'_c|(s-M_N^2)}
\end{equation}
Also, the skewness $\xi$ can be defined as
\begin{equation}
    \xi=-\frac{\Delta\cdot q}{2\bar{P}\cdot q}=\frac{t-M_X^2}{2M_N^2+M_X^2-2s-t}
\end{equation}
where $\bar{P}^\mu=(P+P')^\mu/2$.

In the threshold limit $\sqrt{s}\rightarrow M_N+M_X$, the momentum transfer $t$ is  constrained in the vicinity of $t_{th}=-{M_NM_X^2}/{(M_N+M_X)}$. The kinematically allowed regions are shown on the $(W,-t)$ plane in Fig.\ref{fig:WtheD} for the deuteron  and in Fig.\ref{fig:Wthelium4} for Helium-4.
In the near threshold region $s\gtrsim (M_N+M_X)^2$, the factorization for light nuclei  works when the outgoing meson is heavy enough, so that the  target moves fast enough to be factorized using partons. In the heavy limit, the incoming and outgoing light nuclei velocity is of order 1 up to some correction proportional to the mass ratio $M_N^2/M_X^2$. Hence, the factorization scheme for the parton picture is still satisfied near the threshold of photoproduction. On the other hand, near the threshold region, there is not much energy left to move the heavy meson. The outgoing meson velocity becomes non-relativistic. Therefore, the meson part can be treated using non-relativistic QCD (NRQCD). The skewness $\xi$ near threshold is close to 1. Similar arguments for the photoproduction of heavy mesons on a nucleon near threshold have been used in~\cite{Ma:2003py,Guo:2021ibg,Sun:2021pyw}.

\section{GFFs in the Breit frame}
\label{GFFBreit}
The GFFs in the Breit frame are defined as~\cite{Polyakov:2019lbq}
\begin{widetext}
\bea
\label{eq:GFFBre}
\langle P',\sigma'|T_g^{00}|P,\sigma\rangle&=&2m_D^2\mathcal A^g(t)\delta_{\sigma'\sigma}+\mathcal Q^D (t)\Delta_\alpha \Delta_\beta \langle \sigma'|Q^{\alpha\beta}|\sigma\rangle,
\nonumber\\
\langle P',\sigma'|T_g^{0j}|P,\sigma\rangle
&=&\mathcal J^g(t)m_D\langle \sigma'|(\vec{S}\times i\vec{\Delta})^j|\sigma\rangle ,
\nonumber\\
\langle P',\sigma'|T_g^{jl}|P,\sigma\rangle&=&\mathcal D_0^g(t)\frac{\Delta^j\Delta^l-\delta^{jl}\vec{\Delta}^2}{2}\delta_{m'm}
+\mathcal D_3^g(t)\frac{(\Delta^j\Delta^l-\delta^{jl}\vec{\Delta}^2)\hat \Delta_\alpha \hat \Delta_\beta \langle \sigma'|Q^{\alpha\beta}|\sigma\rangle}{2}
\nonumber\\
&+&\mathcal D_2^D(t)\langle \sigma'|(\Delta^j\Delta^\alpha Q^{l\alpha}+\Delta^l\Delta^\alpha Q^{j\alpha}-\vec{\Delta}^2Q^{jl}-\delta^{jl}Q^{\alpha\beta}\Delta_\alpha \Delta_\beta)|\sigma\rangle .
\eea
\end{widetext}

\bibliography{ref}

\begin{thebibliography}{33}%
\makeatletter
\providecommand \@ifxundefined [1]{%
 \@ifx{#1\undefined}
}%
\providecommand \@ifnum [1]{%
 \ifnum #1\expandafter \@firstoftwo
 \else \expandafter \@secondoftwo
 \fi
}%
\providecommand \@ifx [1]{%
 \ifx #1\expandafter \@firstoftwo
 \else \expandafter \@secondoftwo
 \fi
}%
\providecommand \natexlab [1]{#1}%
\providecommand \enquote  [1]{``#1''}%
\providecommand \bibnamefont  [1]{#1}%
\providecommand \bibfnamefont [1]{#1}%
\providecommand \citenamefont [1]{#1}%
\providecommand \href@noop [0]{\@secondoftwo}%
\providecommand \href [0]{\begingroup \@sanitize@url \@href}%
\providecommand \@href[1]{\@@startlink{#1}\@@href}%
\providecommand \@@href[1]{\endgroup#1\@@endlink}%
\providecommand \@sanitize@url [0]{\catcode `\\12\catcode `\$12\catcode
  `\&12\catcode `\#12\catcode `\^12\catcode `\_12\catcode `\%12\relax}%
\providecommand \@@startlink[1]{}%
\providecommand \@@endlink[0]{}%
\providecommand \url  [0]{\begingroup\@sanitize@url \@url }%
\providecommand \@url [1]{\endgroup\@href {#1}{\urlprefix }}%
\providecommand \urlprefix  [0]{URL }%
\providecommand \Eprint [0]{\href }%
\providecommand \doibase [0]{http://dx.doi.org/}%
\providecommand \selectlanguage [0]{\@gobble}%
\providecommand \bibinfo  [0]{\@secondoftwo}%
\providecommand \bibfield  [0]{\@secondoftwo}%
\providecommand \translation [1]{[#1]}%
\providecommand \BibitemOpen [0]{}%
\providecommand \bibitemStop [0]{}%
\providecommand \bibitemNoStop [0]{.\EOS\space}%
\providecommand \EOS [0]{\spacefactor3000\relax}%
\providecommand \BibitemShut  [1]{\csname bibitem#1\endcsname}%
\let\auto@bib@innerbib\@empty
%</preamble>
\bibitem [{\citenamefont {Sch\"afer}\ and\ \citenamefont
  {Shuryak}(1998)}]{Schafer:1996wv}%
  \BibitemOpen
  \bibfield  {author} {\bibinfo {author} {\bibfnamefont {T.}~\bibnamefont
  {Sch\"afer}}\ and\ \bibinfo {author} {\bibfnamefont {E.~V.}\ \bibnamefont
  {Shuryak}},\ }\href {\doibase 10.1103/RevModPhys.70.323} {\bibfield
  {journal} {\bibinfo  {journal} {Rev. Mod. Phys.}\ }\textbf {\bibinfo {volume}
  {70}},\ \bibinfo {pages} {323} (\bibinfo {year} {1998})},\ \Eprint
  {http://arxiv.org/abs/hep-ph/9610451} {arXiv:hep-ph/9610451} \BibitemShut
  {NoStop}%
\bibitem [{\citenamefont {Biddle}\ \emph {et~al.}(2020)\citenamefont {Biddle},
  \citenamefont {Kamleh},\ and\ \citenamefont {Leinweber}}]{Biddle:2019gke}%
  \BibitemOpen
  \bibfield  {author} {\bibinfo {author} {\bibfnamefont {J.~C.}\ \bibnamefont
  {Biddle}}, \bibinfo {author} {\bibfnamefont {W.}~\bibnamefont {Kamleh}}, \
  and\ \bibinfo {author} {\bibfnamefont {D.~B.}\ \bibnamefont {Leinweber}},\
  }\href {\doibase 10.1103/PhysRevD.102.034504} {\bibfield  {journal} {\bibinfo
   {journal} {Phys. Rev. D}\ }\textbf {\bibinfo {volume} {102}},\ \bibinfo
  {pages} {034504} (\bibinfo {year} {2020})},\ \Eprint
  {http://arxiv.org/abs/1912.09531} {arXiv:1912.09531 [hep-lat]} \BibitemShut
  {NoStop}%
\bibitem [{\citenamefont {Zahed}(2021)}]{Zahed:2021fxk}%
  \BibitemOpen
  \bibfield  {author} {\bibinfo {author} {\bibfnamefont {I.}~\bibnamefont
  {Zahed}},\ }\href {\doibase 10.1103/PhysRevD.104.054031} {\bibfield
  {journal} {\bibinfo  {journal} {Phys. Rev. D}\ }\textbf {\bibinfo {volume}
  {104}},\ \bibinfo {pages} {054031} (\bibinfo {year} {2021})},\ \Eprint
  {http://arxiv.org/abs/2102.08191} {arXiv:2102.08191 [hep-ph]} \BibitemShut
  {NoStop}%
\bibitem [{\citenamefont {Ali}\ \emph {et~al.}(2019)\citenamefont {Ali} \emph
  {et~al.}}]{GlueX:2019mkq}%
  \BibitemOpen
  \bibfield  {author} {\bibinfo {author} {\bibfnamefont {A.}~\bibnamefont
  {Ali}} \emph {et~al.} (\bibinfo {collaboration} {GlueX}),\ }\href {\doibase
  10.1103/PhysRevLett.123.072001} {\bibfield  {journal} {\bibinfo  {journal}
  {Phys. Rev. Lett.}\ }\textbf {\bibinfo {volume} {123}},\ \bibinfo {pages}
  {072001} (\bibinfo {year} {2019})},\ \Eprint
  {http://arxiv.org/abs/1905.10811} {arXiv:1905.10811 [nucl-ex]} \BibitemShut
  {NoStop}%
\bibitem [{\citenamefont {Meziani}\ and\ \citenamefont
  {Joosten}(2020)}]{Meziani:2020oks}%
  \BibitemOpen
  \bibfield  {author} {\bibinfo {author} {\bibfnamefont {Z.-E.}\ \bibnamefont
  {Meziani}}\ and\ \bibinfo {author} {\bibfnamefont {S.}~\bibnamefont
  {Joosten}},\ }in\ \href {\doibase 10.1142/9789811214950_0048} {\emph
  {\bibinfo {booktitle} {{Probing Nucleons and Nuclei in High Energy
  Collisions}: {Dedicated to the Physics of the Electron Ion Collider}}}}\
  (\bibinfo {year} {2020})\ pp.\ \bibinfo {pages} {234--237}\BibitemShut
  {NoStop}%
\bibitem [{\citenamefont {Duran}\ \emph {et~al.}(2023)\citenamefont {Duran}
  \emph {et~al.}}]{Duran:2022xag}%
  \BibitemOpen
  \bibfield  {author} {\bibinfo {author} {\bibfnamefont {B.}~\bibnamefont
  {Duran}} \emph {et~al.},\ }\href {\doibase 10.1038/s41586-023-05730-4}
  {\bibfield  {journal} {\bibinfo  {journal} {Nature}\ }\textbf {\bibinfo
  {volume} {615}},\ \bibinfo {pages} {813} (\bibinfo {year} {2023})},\ \Eprint
  {http://arxiv.org/abs/2207.05212} {arXiv:2207.05212 [nucl-ex]} \BibitemShut
  {NoStop}%
\bibitem [{\citenamefont {Burkert}\ \emph {et~al.}(2023)\citenamefont {Burkert}
  \emph {et~al.}}]{Burkert:2022hjz}%
  \BibitemOpen
  \bibfield  {author} {\bibinfo {author} {\bibfnamefont {V.~D.}\ \bibnamefont
  {Burkert}} \emph {et~al.},\ }\href {\doibase 10.1016/j.ppnp.2023.104032}
  {\bibfield  {journal} {\bibinfo  {journal} {Prog. Part. Nucl. Phys.}\
  }\textbf {\bibinfo {volume} {131}},\ \bibinfo {pages} {104032} (\bibinfo
  {year} {2023})},\ \Eprint {http://arxiv.org/abs/2211.15746} {arXiv:2211.15746
  [nucl-ex]} \BibitemShut {NoStop}%
\bibitem [{\citenamefont {Hatta}\ and\ \citenamefont
  {Yang}(2018)}]{Hatta:2018ina}%
  \BibitemOpen
  \bibfield  {author} {\bibinfo {author} {\bibfnamefont {Y.}~\bibnamefont
  {Hatta}}\ and\ \bibinfo {author} {\bibfnamefont {D.-L.}\ \bibnamefont
  {Yang}},\ }\href {\doibase 10.1103/PhysRevD.98.074003} {\bibfield  {journal}
  {\bibinfo  {journal} {Phys. Rev. D}\ }\textbf {\bibinfo {volume} {98}},\
  \bibinfo {pages} {074003} (\bibinfo {year} {2018})},\ \Eprint
  {http://arxiv.org/abs/1808.02163} {arXiv:1808.02163 [hep-ph]} \BibitemShut
  {NoStop}%
\bibitem [{\citenamefont {Mamo}\ and\ \citenamefont
  {Zahed}(2020)}]{Mamo:2019mka}%
  \BibitemOpen
  \bibfield  {author} {\bibinfo {author} {\bibfnamefont {K.~A.}\ \bibnamefont
  {Mamo}}\ and\ \bibinfo {author} {\bibfnamefont {I.}~\bibnamefont {Zahed}},\
  }\href {\doibase 10.1103/PhysRevD.101.086003} {\bibfield  {journal} {\bibinfo
   {journal} {Phys. Rev. D}\ }\textbf {\bibinfo {volume} {101}},\ \bibinfo
  {pages} {086003} (\bibinfo {year} {2020})},\ \Eprint
  {http://arxiv.org/abs/1910.04707} {arXiv:1910.04707 [hep-ph]} \BibitemShut
  {NoStop}%
\bibitem [{\citenamefont {Kharzeev}(2021)}]{Kharzeev:2021qkd}%
  \BibitemOpen
  \bibfield  {author} {\bibinfo {author} {\bibfnamefont {D.~E.}\ \bibnamefont
  {Kharzeev}},\ }\href {\doibase 10.1103/PhysRevD.104.054015} {\bibfield
  {journal} {\bibinfo  {journal} {Phys. Rev. D}\ }\textbf {\bibinfo {volume}
  {104}},\ \bibinfo {pages} {054015} (\bibinfo {year} {2021})},\ \Eprint
  {http://arxiv.org/abs/2102.00110} {arXiv:2102.00110 [hep-ph]} \BibitemShut
  {NoStop}%
\bibitem [{\citenamefont {Ji}(2021)}]{Ji:2021mtz}%
  \BibitemOpen
  \bibfield  {author} {\bibinfo {author} {\bibfnamefont {X.}~\bibnamefont
  {Ji}},\ }\href {\doibase 10.1007/s11467-021-1065-x} {\bibfield  {journal}
  {\bibinfo  {journal} {Front. Phys. (Beijing)}\ }\textbf {\bibinfo {volume}
  {16}},\ \bibinfo {pages} {64601} (\bibinfo {year} {2021})},\ \Eprint
  {http://arxiv.org/abs/2102.07830} {arXiv:2102.07830 [hep-ph]} \BibitemShut
  {NoStop}%
\bibitem [{\citenamefont {Hatta}\ and\ \citenamefont
  {Strikman}(2021)}]{Hatta:2021can}%
  \BibitemOpen
  \bibfield  {author} {\bibinfo {author} {\bibfnamefont {Y.}~\bibnamefont
  {Hatta}}\ and\ \bibinfo {author} {\bibfnamefont {M.}~\bibnamefont
  {Strikman}},\ }\href {\doibase 10.1016/j.physletb.2021.136295} {\bibfield
  {journal} {\bibinfo  {journal} {Phys. Lett. B}\ }\textbf {\bibinfo {volume}
  {817}},\ \bibinfo {pages} {136295} (\bibinfo {year} {2021})},\ \Eprint
  {http://arxiv.org/abs/2102.12631} {arXiv:2102.12631 [hep-ph]} \BibitemShut
  {NoStop}%
\bibitem [{\citenamefont {Guo}\ \emph {et~al.}(2021)\citenamefont {Guo},
  \citenamefont {Ji},\ and\ \citenamefont {Liu}}]{Guo:2021ibg}%
  \BibitemOpen
  \bibfield  {author} {\bibinfo {author} {\bibfnamefont {Y.}~\bibnamefont
  {Guo}}, \bibinfo {author} {\bibfnamefont {X.}~\bibnamefont {Ji}}, \ and\
  \bibinfo {author} {\bibfnamefont {Y.}~\bibnamefont {Liu}},\ }\href {\doibase
  10.1103/PhysRevD.103.096010} {\bibfield  {journal} {\bibinfo  {journal}
  {Phys. Rev. D}\ }\textbf {\bibinfo {volume} {103}},\ \bibinfo {pages}
  {096010} (\bibinfo {year} {2021})},\ \Eprint
  {http://arxiv.org/abs/2103.11506} {arXiv:2103.11506 [hep-ph]} \BibitemShut
  {NoStop}%
\bibitem [{\citenamefont {Sun}\ \emph {et~al.}(2021)\citenamefont {Sun},
  \citenamefont {Tong},\ and\ \citenamefont {Yuan}}]{Sun:2021gmi}%
  \BibitemOpen
  \bibfield  {author} {\bibinfo {author} {\bibfnamefont {P.}~\bibnamefont
  {Sun}}, \bibinfo {author} {\bibfnamefont {X.-B.}\ \bibnamefont {Tong}}, \
  and\ \bibinfo {author} {\bibfnamefont {F.}~\bibnamefont {Yuan}},\ }\href
  {\doibase 10.1016/j.physletb.2021.136655} {\bibfield  {journal} {\bibinfo
  {journal} {Phys. Lett. B}\ }\textbf {\bibinfo {volume} {822}},\ \bibinfo
  {pages} {136655} (\bibinfo {year} {2021})},\ \Eprint
  {http://arxiv.org/abs/2103.12047} {arXiv:2103.12047 [hep-ph]} \BibitemShut
  {NoStop}%
\bibitem [{\citenamefont {Mamo}\ and\ \citenamefont
  {Zahed}(2022)}]{Mamo:2022eui}%
  \BibitemOpen
  \bibfield  {author} {\bibinfo {author} {\bibfnamefont {K.~A.}\ \bibnamefont
  {Mamo}}\ and\ \bibinfo {author} {\bibfnamefont {I.}~\bibnamefont {Zahed}},\
  }\href {\doibase 10.1103/PhysRevD.106.086004} {\bibfield  {journal} {\bibinfo
   {journal} {Phys. Rev. D}\ }\textbf {\bibinfo {volume} {106}},\ \bibinfo
  {pages} {086004} (\bibinfo {year} {2022})},\ \Eprint
  {http://arxiv.org/abs/2204.08857} {arXiv:2204.08857 [hep-ph]} \BibitemShut
  {NoStop}%
\bibitem [{\citenamefont {Wang}\ \emph {et~al.}(2022)\citenamefont {Wang},
  \citenamefont {Zeng},\ and\ \citenamefont {Wang}}]{Wang:2022vhr}%
  \BibitemOpen
  \bibfield  {author} {\bibinfo {author} {\bibfnamefont {X.-Y.}\ \bibnamefont
  {Wang}}, \bibinfo {author} {\bibfnamefont {F.}~\bibnamefont {Zeng}}, \ and\
  \bibinfo {author} {\bibfnamefont {Q.}~\bibnamefont {Wang}},\ }\href {\doibase
  10.1103/PhysRevD.105.096033} {\bibfield  {journal} {\bibinfo  {journal}
  {Phys. Rev. D}\ }\textbf {\bibinfo {volume} {105}},\ \bibinfo {pages}
  {096033} (\bibinfo {year} {2022})},\ \Eprint
  {http://arxiv.org/abs/2204.07294} {arXiv:2204.07294 [hep-ph]} \BibitemShut
  {NoStop}%
\bibitem [{\citenamefont {Guo}\ \emph {et~al.}(2023)\citenamefont {Guo},
  \citenamefont {Ji}, \citenamefont {Liu},\ and\ \citenamefont
  {Yang}}]{Guo:2023pqw}%
  \BibitemOpen
  \bibfield  {author} {\bibinfo {author} {\bibfnamefont {Y.}~\bibnamefont
  {Guo}}, \bibinfo {author} {\bibfnamefont {X.}~\bibnamefont {Ji}}, \bibinfo
  {author} {\bibfnamefont {Y.}~\bibnamefont {Liu}}, \ and\ \bibinfo {author}
  {\bibfnamefont {J.}~\bibnamefont {Yang}},\ }\href {\doibase
  10.1103/PhysRevD.108.034003} {\bibfield  {journal} {\bibinfo  {journal}
  {Phys. Rev. D}\ }\textbf {\bibinfo {volume} {108}},\ \bibinfo {pages}
  {034003} (\bibinfo {year} {2023})},\ \Eprint
  {http://arxiv.org/abs/2305.06992} {arXiv:2305.06992 [hep-ph]} \BibitemShut
  {NoStop}%
\bibitem [{\citenamefont {Nemchik}\ \emph {et~al.}(1997)\citenamefont
  {Nemchik}, \citenamefont {Nikolaev}, \citenamefont {Predazzi},\ and\
  \citenamefont {Zakharov}}]{Nemchik:1996cw}%
  \BibitemOpen
  \bibfield  {author} {\bibinfo {author} {\bibfnamefont {J.}~\bibnamefont
  {Nemchik}}, \bibinfo {author} {\bibfnamefont {N.~N.}\ \bibnamefont
  {Nikolaev}}, \bibinfo {author} {\bibfnamefont {E.}~\bibnamefont {Predazzi}},
  \ and\ \bibinfo {author} {\bibfnamefont {B.~G.}\ \bibnamefont {Zakharov}},\
  }\href {\doibase 10.1007/s002880050448} {\bibfield  {journal} {\bibinfo
  {journal} {Z. Phys. C}\ }\textbf {\bibinfo {volume} {75}},\ \bibinfo {pages}
  {71} (\bibinfo {year} {1997})},\ \Eprint
  {http://arxiv.org/abs/hep-ph/9605231} {arXiv:hep-ph/9605231} \BibitemShut
  {NoStop}%
\bibitem [{\citenamefont {Abazov}\ \emph {et~al.}(2021)\citenamefont {Abazov}
  \emph {et~al.}}]{TOTEM:2020zzr}%
  \BibitemOpen
  \bibfield  {author} {\bibinfo {author} {\bibfnamefont {V.~M.}\ \bibnamefont
  {Abazov}} \emph {et~al.} (\bibinfo {collaboration} {TOTEM, D0}),\ }\href
  {\doibase 10.1103/PhysRevLett.127.062003} {\bibfield  {journal} {\bibinfo
  {journal} {Phys. Rev. Lett.}\ }\textbf {\bibinfo {volume} {127}},\ \bibinfo
  {pages} {062003} (\bibinfo {year} {2021})},\ \Eprint
  {http://arxiv.org/abs/2012.03981} {arXiv:2012.03981 [hep-ex]} \BibitemShut
  {NoStop}%
\bibitem [{\citenamefont {Antchev}\ \emph {et~al.}(2019)\citenamefont {Antchev}
  \emph {et~al.}}]{TOTEM:2018hki}%
  \BibitemOpen
  \bibfield  {author} {\bibinfo {author} {\bibfnamefont {G.}~\bibnamefont
  {Antchev}} \emph {et~al.} (\bibinfo {collaboration} {TOTEM}),\ }\href
  {\doibase 10.1140/epjc/s10052-019-7346-7} {\bibfield  {journal} {\bibinfo
  {journal} {Eur. Phys. J. C}\ }\textbf {\bibinfo {volume} {79}},\ \bibinfo
  {pages} {861} (\bibinfo {year} {2019})},\ \Eprint
  {http://arxiv.org/abs/1812.08283} {arXiv:1812.08283 [hep-ex]} \BibitemShut
  {NoStop}%
\bibitem [{\citenamefont {Mamo}\ and\ \citenamefont
  {Zahed}(2021)}]{Mamo:2021krl}%
  \BibitemOpen
  \bibfield  {author} {\bibinfo {author} {\bibfnamefont {K.~A.}\ \bibnamefont
  {Mamo}}\ and\ \bibinfo {author} {\bibfnamefont {I.}~\bibnamefont {Zahed}},\
  }\href {\doibase 10.1103/PhysRevD.103.094010} {\bibfield  {journal} {\bibinfo
   {journal} {Phys. Rev. D}\ }\textbf {\bibinfo {volume} {103}},\ \bibinfo
  {pages} {094010} (\bibinfo {year} {2021})},\ \Eprint
  {http://arxiv.org/abs/2103.03186} {arXiv:2103.03186 [hep-ph]} \BibitemShut
  {NoStop}%
\bibitem [{\citenamefont {Abidin}\ and\ \citenamefont
  {Carlson}(2009)}]{Abidin:2009hr}%
  \BibitemOpen
  \bibfield  {author} {\bibinfo {author} {\bibfnamefont {Z.}~\bibnamefont
  {Abidin}}\ and\ \bibinfo {author} {\bibfnamefont {C.~E.}\ \bibnamefont
  {Carlson}},\ }\href {\doibase 10.1103/PhysRevD.79.115003} {\bibfield
  {journal} {\bibinfo  {journal} {Phys. Rev. D}\ }\textbf {\bibinfo {volume}
  {79}},\ \bibinfo {pages} {115003} (\bibinfo {year} {2009})},\ \Eprint
  {http://arxiv.org/abs/0903.4818} {arXiv:0903.4818 [hep-ph]} \BibitemShut
  {NoStop}%
\bibitem [{\citenamefont {Guzey}\ \emph {et~al.}(2022)\citenamefont {Guzey},
  \citenamefont {Rinaldi}, \citenamefont {Scopetta}, \citenamefont {Strikman},\
  and\ \citenamefont {Viviani}}]{Guzey:2022jtv}%
  \BibitemOpen
  \bibfield  {author} {\bibinfo {author} {\bibfnamefont {V.}~\bibnamefont
  {Guzey}}, \bibinfo {author} {\bibfnamefont {M.}~\bibnamefont {Rinaldi}},
  \bibinfo {author} {\bibfnamefont {S.}~\bibnamefont {Scopetta}}, \bibinfo
  {author} {\bibfnamefont {M.}~\bibnamefont {Strikman}}, \ and\ \bibinfo
  {author} {\bibfnamefont {M.}~\bibnamefont {Viviani}},\ }\href {\doibase
  10.1103/PhysRevLett.129.242503} {\bibfield  {journal} {\bibinfo  {journal}
  {Phys. Rev. Lett.}\ }\textbf {\bibinfo {volume} {129}},\ \bibinfo {pages}
  {242503} (\bibinfo {year} {2022})},\ \Eprint
  {http://arxiv.org/abs/2202.12200} {arXiv:2202.12200 [hep-ph]} \BibitemShut
  {NoStop}%
\bibitem [{\citenamefont {Liu}\ and\ \citenamefont
  {Zahed}(2024)}]{Liu:2024yqa}%
  \BibitemOpen
  \bibfield  {author} {\bibinfo {author} {\bibfnamefont {W.-Y.}\ \bibnamefont
  {Liu}}\ and\ \bibinfo {author} {\bibfnamefont {I.}~\bibnamefont {Zahed}},\
  }\href@noop {} {\  (\bibinfo {year} {2024})},\ \Eprint
  {http://arxiv.org/abs/2404.03875} {arXiv:2404.03875 [hep-ph]} \BibitemShut
  {NoStop}%
\bibitem [{\citenamefont {He}\ and\ \citenamefont
  {Zahed}(2024{\natexlab{a}})}]{He:2023ogg}%
  \BibitemOpen
  \bibfield  {author} {\bibinfo {author} {\bibfnamefont {F.}~\bibnamefont
  {He}}\ and\ \bibinfo {author} {\bibfnamefont {I.}~\bibnamefont {Zahed}},\
  }\href {\doibase 10.1103/PhysRevC.109.045209} {\bibfield  {journal} {\bibinfo
   {journal} {Phys. Rev. C}\ }\textbf {\bibinfo {volume} {109}},\ \bibinfo
  {pages} {045209} (\bibinfo {year} {2024}{\natexlab{a}})},\ \Eprint
  {http://arxiv.org/abs/2310.12315} {arXiv:2310.12315 [nucl-th]} \BibitemShut
  {NoStop}%
\bibitem [{\citenamefont {He}\ and\ \citenamefont
  {Zahed}(2024{\natexlab{b}})}]{He:2024vzz}%
  \BibitemOpen
  \bibfield  {author} {\bibinfo {author} {\bibfnamefont {F.}~\bibnamefont
  {He}}\ and\ \bibinfo {author} {\bibfnamefont {I.}~\bibnamefont {Zahed}},\
  }\href {\doibase 10.1103/PhysRevC.110.014312} {\bibfield  {journal} {\bibinfo
   {journal} {Phys. Rev. C}\ }\textbf {\bibinfo {volume} {110}},\ \bibinfo
  {pages} {014312} (\bibinfo {year} {2024}{\natexlab{b}})},\ \Eprint
  {http://arxiv.org/abs/2401.09318} {arXiv:2401.09318 [nucl-th]} \BibitemShut
  {NoStop}%
\bibitem [{\citenamefont {He}\ and\ \citenamefont
  {Zahed}(2024{\natexlab{c}})}]{He:2024jgc}%
  \BibitemOpen
  \bibfield  {author} {\bibinfo {author} {\bibfnamefont {F.}~\bibnamefont
  {He}}\ and\ \bibinfo {author} {\bibfnamefont {I.}~\bibnamefont {Zahed}},\
  }\href@noop {} {\  (\bibinfo {year} {2024}{\natexlab{c}})},\ \Eprint
  {http://arxiv.org/abs/2406.07412} {arXiv:2406.07412 [nucl-th]} \BibitemShut
  {NoStop}%
\bibitem [{\citenamefont {Sun}\ \emph {et~al.}(2022)\citenamefont {Sun},
  \citenamefont {Tong},\ and\ \citenamefont {Yuan}}]{Sun:2021pyw}%
  \BibitemOpen
  \bibfield  {author} {\bibinfo {author} {\bibfnamefont {P.}~\bibnamefont
  {Sun}}, \bibinfo {author} {\bibfnamefont {X.-B.}\ \bibnamefont {Tong}}, \
  and\ \bibinfo {author} {\bibfnamefont {F.}~\bibnamefont {Yuan}},\ }\href
  {\doibase 10.1103/PhysRevD.105.054032} {\bibfield  {journal} {\bibinfo
  {journal} {Phys. Rev. D}\ }\textbf {\bibinfo {volume} {105}},\ \bibinfo
  {pages} {054032} (\bibinfo {year} {2022})},\ \Eprint
  {http://arxiv.org/abs/2111.07034} {arXiv:2111.07034 [hep-ph]} \BibitemShut
  {NoStop}%
\bibitem [{\citenamefont {Polyakov}\ and\ \citenamefont
  {Sun}(2019)}]{Polyakov:2019lbq}%
  \BibitemOpen
  \bibfield  {author} {\bibinfo {author} {\bibfnamefont {M.~V.}\ \bibnamefont
  {Polyakov}}\ and\ \bibinfo {author} {\bibfnamefont {B.-D.}\ \bibnamefont
  {Sun}},\ }\href {\doibase 10.1103/PhysRevD.100.036003} {\bibfield  {journal}
  {\bibinfo  {journal} {Phys. Rev. D}\ }\textbf {\bibinfo {volume} {100}},\
  \bibinfo {pages} {036003} (\bibinfo {year} {2019})},\ \Eprint
  {http://arxiv.org/abs/1903.02738} {arXiv:1903.02738 [hep-ph]} \BibitemShut
  {NoStop}%
\bibitem [{\citenamefont {Tanabashi}\ \emph {et~al.}(2018)\citenamefont
  {Tanabashi} \emph {et~al.}}]{ParticleDataGroup:2018ovx}%
  \BibitemOpen
  \bibfield  {author} {\bibinfo {author} {\bibfnamefont {M.}~\bibnamefont
  {Tanabashi}} \emph {et~al.} (\bibinfo {collaboration} {Particle Data
  Group}),\ }\href {\doibase 10.1103/PhysRevD.98.030001} {\bibfield  {journal}
  {\bibinfo  {journal} {Phys. Rev. D}\ }\textbf {\bibinfo {volume} {98}},\
  \bibinfo {pages} {030001} (\bibinfo {year} {2018})}\BibitemShut {NoStop}%
\bibitem [{\citenamefont {Bodwin}\ \emph {et~al.}(2006)\citenamefont {Bodwin},
  \citenamefont {Braaten}, \citenamefont {Lee},\ and\ \citenamefont
  {Yu}}]{Bodwin:2006yd}%
  \BibitemOpen
  \bibfield  {author} {\bibinfo {author} {\bibfnamefont {G.~T.}\ \bibnamefont
  {Bodwin}}, \bibinfo {author} {\bibfnamefont {E.}~\bibnamefont {Braaten}},
  \bibinfo {author} {\bibfnamefont {J.}~\bibnamefont {Lee}}, \ and\ \bibinfo
  {author} {\bibfnamefont {C.}~\bibnamefont {Yu}},\ }\href {\doibase
  10.1103/PhysRevD.74.074014} {\bibfield  {journal} {\bibinfo  {journal} {Phys.
  Rev. D}\ }\textbf {\bibinfo {volume} {74}},\ \bibinfo {pages} {074014}
  (\bibinfo {year} {2006})},\ \Eprint {http://arxiv.org/abs/hep-ph/0608200}
  {arXiv:hep-ph/0608200} \BibitemShut {NoStop}%
\bibitem [{\citenamefont {Adhikari}\ \emph {et~al.}(2023)\citenamefont
  {Adhikari} \emph {et~al.}}]{GlueX:2023pev}%
  \BibitemOpen
  \bibfield  {author} {\bibinfo {author} {\bibfnamefont {S.}~\bibnamefont
  {Adhikari}} \emph {et~al.} (\bibinfo {collaboration} {GlueX}),\ }\href
  {\doibase 10.1103/PhysRevC.108.025201} {\bibfield  {journal} {\bibinfo
  {journal} {Phys. Rev. C}\ }\textbf {\bibinfo {volume} {108}},\ \bibinfo
  {pages} {025201} (\bibinfo {year} {2023})},\ \Eprint
  {http://arxiv.org/abs/2304.03845} {arXiv:2304.03845 [nucl-ex]} \BibitemShut
  {NoStop}%
\bibitem [{\citenamefont {Ma}(2003)}]{Ma:2003py}%
  \BibitemOpen
  \bibfield  {author} {\bibinfo {author} {\bibfnamefont {J.~P.}\ \bibnamefont
  {Ma}},\ }\href {\doibase 10.1016/j.nuclphysa.2003.08.016} {\bibfield
  {journal} {\bibinfo  {journal} {Nucl. Phys. A}\ }\textbf {\bibinfo {volume}
  {727}},\ \bibinfo {pages} {333} (\bibinfo {year} {2003})},\ \Eprint
  {http://arxiv.org/abs/hep-ph/0301155} {arXiv:hep-ph/0301155} \BibitemShut
  {NoStop}%
\end{thebibliography}%

\end{document}